\documentstyle{mn}                 %
\def\ni{$^{56}$Ni}
\def\co{$^{56}$Co}
\def\d{SN~1997D}
\def\a{SN~1987A}

\def\ef{SN~1997ef}
\def\bw{SN~1998bw}

\def\kms{km\,s$^{-1}$}
\def\mcento{mag\,(100d)$^{-1}$}

\def\Ha{H$\alpha$}
\def\Hb{H$\beta$}

\def\Mv{M$_{\rm B}$}

\def\M{M$_{\odot}$}

\def\e#1{$\times$ $10^{#1}$ }
\def\ms{$M_\odot$}

\input{psfig.tex}

\title[Supernova 1997D]
{The fading of SN~1997D
\thanks{Based on observations collected at ESO-La Silla (Chile)}}

\author[Benetti et al.]
{S. Benetti$^{1}$, M. Turatto$^2$,
S. Balberg$^{3}$, L. Zampieri$^{3,4}$, S.L. Shapiro$^{3,5}$,
\and E. Cappellaro$^2$, K. Nomoto$^{6,7}$, T. Nakamura$^6$,
\and P.A. Mazzali$^8$, F. Patat$^9$
\\
$1$Telescopio Nazionale Galileo, Apartado de Correos 565, E-38700
Santa Cruz de La Palma, Canary Islands, Spain \\
$^2$Osservatorio Astronomico di Padova, Vicolo dell'Osservatorio 5,
I-35122 Padova, Italy \\
$^3$Department of Physics, Loomis Laboratory of Physics,
University of Illinois at Urbana--Champaign, \\
1110 West Green Street, Urbana, IL 61801--3080 \\
$^4$Dipartimento di Fisica, Universit\`a degli Studi di Padova,
Via Marzolo 8, 35131 Padova, Italy \\
$^5$Department of Astronomy and National Center for Supercomputing
Applications, University of Illinois at Urbana--Champaign, \\
Urbana, IL 61801 \\
$^6$Department of Astronomy, School of Science, 
University of Tokyo, Tokyo, Japan\\
$^7$Research Center for the Early Universe, School of Science, 
University of Tokyo, Tokyo, Japan\\
$^8$Osservatorio Astronomico di Trieste, Via G.B. Tiepolo 11, I-34131
Trieste, Italy\\
$^9$European Southern Observatory, Karl-Schwarzschild-Strasse 2, D-85748
Garching, Germany
}

\date{Received ................; accepted ................}

\begin{document}

\maketitle

\begin{abstract}

We present a new set of spectroscopic and photometric data extending
the observations of SN~1997D to over 400 days after the explosion. These
observations confirm the peculiar properties of SN~1997D, such as the
very low abundance of $^{56}$Co (0.002 \M\/) and the low expansion
velocity of the ejecta ($\sim 1000$ km s$^{-1}$). We discuss the
implications of these observations for the character of the progenitor
and the nature of the remnant showing that a Crab-like pulsar or an
accreting neutron star formed in the explosion of a low mass
progenitor should already have produced a detectable luminosity at
this epoch, in contrast with photometric data. On the other hand, the
explosion of a high mass progenitor with the formation of a black hole
is consistent with the available observations. The consequences of
this conclusion regarding the nature of the explosion and the
prospects of directly identifying the black hole are also addressed.

\end{abstract}


\section{Introduction} \label{int}

SN~1997D is the least luminous and least energetic type II supernova
discovered to date. An earlier analysis (Turatto et al.~1998, hereafter
PaperI) showed that the SN peak magnitude was fainter than
\Mv$=-14.65$, had a very red color, indicating that the
photospheric temperature at discovery was only $T_{eff}=6400$\degr K,
and a very slow expansion velocities of the order of 1000
\kms\/.  The luminosity in the late light curve indicated that the
ejected \ni\ mass was only 0.002\M, at least one order of magnitude
smaller than the estimates for normal type II supernova
(SNII).

A similar small amount of \ni\/ was suggested for another Type
II supernova, SN~1994W \cite{solle}. However Sollerman et
al. (1998) argues that the light curve of SN~1994W is contaminated in
different phases by the probable contribution of CSM-ejecta
interaction and by dust formation in the ejecta which makes this case
less compelling than that of SN~1997D.

The unusual properties make SN~1997D an intriguing target for further
study, especially with regards to the character of the progenitor and
the nature of the compact remnant left after the explosion.  In
the following we will discuss two extreme alternatives: a high mass
(25--40\M\/) star in which the low mass of \ni\/ is the result of the
fallback of material onto the collapsed core
\cite{WoosWeav95}, or a low mass (8--10\M\/) star in which hardly any \ni\/ is
produced to begin with (Nomoto et al.~1982; Nomoto 1984, 1987). In
PaperI, based on modeling of the light curve and spectrum, we favored
a scenario in which a high mass (26\M\/) progenitor exploded about 50
days before the discovery with an explosion energy of only $4 \times
10^{50}$ erg. Instead, Chugai \& Utrobin (2000) using a hydrodynamical
model and an analysis of the nebular spectrum suggested that a low
mass progenitor (and a low explosion energy) is preferred.
Identifying which of the two scenarios lies behind SN~1997D is of
significant interest, as well as being an observational challenge.  In
particular, there would be several intriguing consequences if the
progenitor was indeed a high mass star, which would indicate the
existence of a novel sub-class of low-energy SNII.

Crucial to our discussion is the fact that the two scenarios differ
qualitatively with regard to the nature of the compact remnant they
produce. In the case of a low mass progenitor, the likely remnant is a
neutron star \cite{Nomoto-et-al}, while for a high mass progenitor the
rate of early fallback during the explosion is expected to induce the
collapse of the proto-neutron star into a black hole (Colgate 1971;
Chevalier 1989; Woosley \& Weaver 1995; Woosley \& Timmes 1996; Fryer
1999). The supernova birth of a black hole has recently found an
important indirect observational proof from the detection of an
overabundance of $\alpha$-elements in the atmosphere of the companion
star orbiting a probable black hole in the X-ray binary system GRO
J1655-40 \cite{Israelianetal99}.

A direct observation of the effect of the newly formed black hole on
the light curve is usually impossible, owing to the large abundance of
radioactive isotopes. Zampieri et al.~(1998b) found that if SN~1987A
had produced a black hole, it would emerge as late as 900 years after
the explosion at a luminosity of $\sim
10^{32}\;\mbox{erg}\;\mbox{s}^{-1}$. In the case of
\d\/, Zampieri et al.~(1998a) pointed out that because of the very low absolute
magnitude due to the low abundance of radioactive isotopes in the
ejecta, the contribution by late time accretion of material onto the
black hole could emerge above the radioactive decay about {\it three
years} after the explosion.  Since this luminosity is expected to have
a characteristic decay in time of $t^{-25/18}$ (rather than the
exponential decline typical of radioactive sources), it should leave a
distinguishable imprint in the late time light curve. This estimate
was recently confirmed by a detailed investigation by Balberg et
al.~(2000), who found that the luminosity at emergence should be
$0.5-3\times 10^{36}\;\mbox{erg}\;\mbox{s}^{-1}$. Such a luminosity,
at the distance of SN~1997D, is marginally detectable with HST, so, in
principle, SN~1997D could offer a first {\it direct} observation of a
black hole in a supernova.

In this work, we revisit the unusual character of \d\/ and
present the full set of photometric and spectroscopic data up to
about 400 days after the explosion.
%, which extends the early observations presented in PaperI. 
In \S~3 we discuss the decline of
the late time light curve in connection with the nature of the
progenitor of \d\/ and its compact remnant and show that the presence
of a black hole is consistent with the available data. We present our
conclusions and discuss several implications that arise from them in
\S~4. In particular, we emphasize the importance of extending the
observations of SN~1997D to the detection limit of the available ground
and space based instruments.

\section{The available data set} \label{obs}

\subsection{Photometry} \label{phot}

The main data relative to
\d\/ and to the parent galaxy are summarized in Tab. \ref{data} (cfr. PaperI).

\begin{table}
\caption{Main data of SN~1997D and NGC~1536}\label{data}
\begin{tabular}{ll}
\hline
Parent galaxy & NGC 1536 \\ Galaxy type & SB(s)c pec: $^\dag$ \\
v$_{helio}$ & 1296 \kms $^\dag$ \\ Distance modulus (H$_0=75$) &
30.64$^{\ddag}$ \\ Galactic A$_B$ & 0.0$^{\ddag}$ \\ & \\ SN type & II
peculiar \\ magnitude at max & $V_{max} < 16$ \\ RA(SN) (2000) &
$04^h11^m01.0^s$ $^*$ \\ Dec(SN) (2000) & $-56$\degr29\arcmin56\farcs0
$^*$ \\ Offset from nucleus & 11\arcsec~E ~~ 43\arcsec~S $^*$ \\ Date
of discovery & 1997 Jan. 14.15 UT \\ Date of maximum & JD $\sim
2450430 \pm20$ (1996 Dec. 12) \\ L$_{bol}$ at maximum & $\ge
1.23\times 10^{41}$ erg s$^{-1}$\\

\hline
\end{tabular}

{\dag} RC3; {\ddag} Tully (1988); $^*$ Barbon et al. 1999

\end{table}

All the photometric measurements available to date are listed in
Tab.~\ref{obs_tab}. Observations were obtained on 27 different epochs
up to about 400 days after explosion. Data reduction followed the
standard procedures making use for the SN measurements, of a PSF
fitting technique. The mean photometric errors, estimated with
artificial stars experiments, are about 0.05 mag when the SN was
brighter than $\sim 19.5$ mag, 0.10 mag for magnitudes ranging between
19.5$-$21 and 0.2 mag for the faintest magnitudes.

\begin{figure}
\psfig{figure=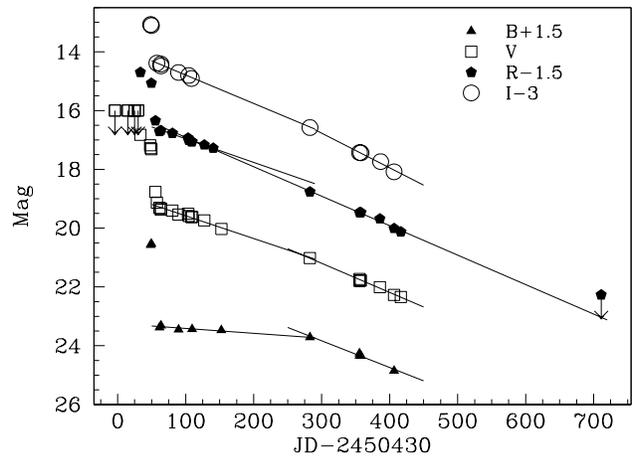,width=9cm,angle=270}
\caption{B, V, R and I light curves of SN~1997D. Three of the prediscovery
upper limits reported in PaperI are shown.}
\label{phot_fig}
\end{figure}

\begin{figure}
\psfig{figure=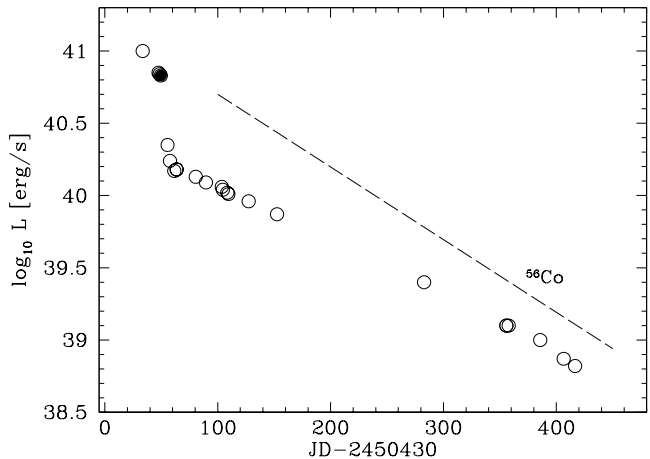,width=9cm,angle=270}
\caption{Bolometric light curve of SN~1997D. The filled circle includes all the
flux between 0.35 and 2.5$\mu$. Open circles represent the BVRI flux
scaled to match the filled circle at the common epoch. The $^{56}$Co
decay line is also reported for comparison.}
\label{bol_fig}
\end{figure}

The B, V, R and I light curves of SN~1997D are shown in
Fig.~\ref{phot_fig}. On the basis of the light curve modeling, in
PaperI we argued that the maximum occurred around J.D. 2450430 ($\pm
20$d) and that the SN has been caught at the end of the plateau phase
on the way to reach the radioactive tail. With the new data the late
time light curve appears to show two distinct behaviors: from day 70
to about 200 the decline rate ${\gamma}$ is much steeper at red
wavelengths than at blue ones (${\gamma}_{\rm B}=0.16$ \mcento,
${\gamma}_{\rm V}=0.78$ \mcento, ${\gamma}_{\rm R}=0.80$ \mcento and
${\gamma}_{\rm I}=0.96$ \mcento), while after day $\sim$200 the
slopes are similar to each other (${\gamma}_{\rm B}=0.91$ \mcento,
${\gamma}_{\rm V}=0.99$ \mcento, ${\gamma}_{\rm R}=1.00$ \mcento,
${\gamma}_{\rm I}= 1.18$ \mcento)and remarkably close to that of \co\/
(${\gamma}=0.98$ \mcento).

We used the multicolor photometry to estimate the bolometric light
curve shown in Fig.~\ref{bol_fig}. This has been obtained by
integrating the flux in the BVRI bands, and then scaling up the curve
to match the spectrophotometric observation of Jan. 27-31 (filled
symbol in Fig.~\ref{bol_fig}) spanning the range 0.35 to 2.5
$\mu$. This means we made the assumption that the IR contribution to
the total luminosity remains constant at later epochs, consistently
with the finding of Schmidt \shortcite{schmidt}. We note that most of
the flux is emitted in the V, R and I bands and that, owing to the
slow decline, the contribution of the B band becomes significant only
after 250 days. Looking in more detail we find that the decline of the
bolometric luminosity in the period 60--200 days, $ 0.89\pm 0.02$
\mcento, is marginally slower and instead, after day 200, ${\gamma}_{\rm
bol}=1.07\pm 0.03$ \mcento, is slightly higher than that
expected if the energy is supplied by the decay of \co\/.

The ratio of the bolometric luminosities of the last segment of the
light curve with the corresponding one of \a\/ is
$L(87A)/L(97D)=32.5$, somewhat smaller than the value derived
early on (PaperI). Assuming for \d\/ the same $\gamma-$deposition as
in \a, we confirm our earlier suggestion that the mass of
ejected \ni\/ is about 0.002 \M.

\begin{table}
\caption{Photometry of SN~1997D}\label{obs_tab}
\begin{flushleft}
\begin{tabular}{lcccccl}
\hline
    date & J.D.     & B    & V    &  R    &  I   &instr\\
         & 2400000+ &      &      &       &      &\\
\hline
15/01/97 & 50463.52 &      & 16.82& 16.20&      & Dut\\
29/01/97 & 50477.62 &      & 17.18&      &      & CT91\\
30/01/97 & 50478.56 &19.07 & 17.28&      & 16.07& CT91\\
31/01/97 & 50479.58 &19.03 & 17.32& 16.57& 16.11& Dut(*)\\
06/02/97 & 50485.56 &       & 18.77& 17.84&      & 3.6\\
08/02/97 & 50487.68 &       & 19.13&      & 17.39& CT91\\
12/02/97 & 50491.54 &21.88 & 19.31& 18.21&      & 2.2\\
14/02/97 & 50493.60 &21.82 & 19.38& 18.18& 17.42& 2.2\\
14/02/97 & 50493.60 &$\le21.6$& 19.33&   & 17.48& CT91\\
03/03/97 & 50510.52 &      & 19.40& 18.27&      & 3.6\\
12/03/97 & 50519.59 &21.95 & 19.54&      & 17.71& CT91\\
26/03/97 & 50533.60 &      & 19.51& 18.44&      & NTT\\
27/03/97 & 50534.51 &      & 19.59& 18.51& 17.81& CT91\\
31/03/97 & 50538.52 &      & 19.62& 18.56& 17.91& CT91\\
01/04/97 & 50539.52 &21.94 & 19.62& 18.56&      & Dut\\
18/04/97 & 50557.48 &      & 19.74& 18.67&      & Dut\\
01/05/97 & 50570.47 &       &      & 18.78&     & 3.6\\
13/05/97 & 50582.49 &21.97 & 20.03&      &      & 2.2\\
21/09/97 & 50712.80 &22.21 & 21.02& 20.27& 19.58& 2.2\\
03/12/97 & 50785.75 &22.84 & 21.80& 20.97& 20.43& Dan\\
03/12/97 & 50785.75 &22.76 & 21.73&      &      & Dan\\
05/12/97 & 50787.80 &22.84 & 21.78& 20.97& 20.44& Dan\\
02/01/98 & 50815.64 &      & 22.01& 21.18&      & Dut\\
03/01/98 & 50816.75 &      &      &      & 20.74& Dut\\
23/01/98 & 50836.50 &23.35 & 22.27& 21.51& 21.08& Dan\\
02/02/98 & 50846.56 &      & 22.34& 21.63&      & EF2\\
23/11/98 & 51141.15 &      &      & $\le 23.8$& & EF2\\
\hline
\end{tabular}

($^*$)~U=21.33

Dut = Dutch 0.9m+CCD Cam.; CT91 = CTIO 0.91m+CCD Cam.\\
3.6 = ESO 3.6m+EFOSC1; 2.2 = MPI/ESO 2.2m+EFOSC2\\
NTT = ESO NTT+SUSI; Dan = Danish 1.54m+DFOSC\\
EF2 = ESO 3.6m+EFOSC2

\end{flushleft}
\end{table}

\subsection{Spectroscopy} \label{spec} 

\begin{figure*}
\psfig{figure=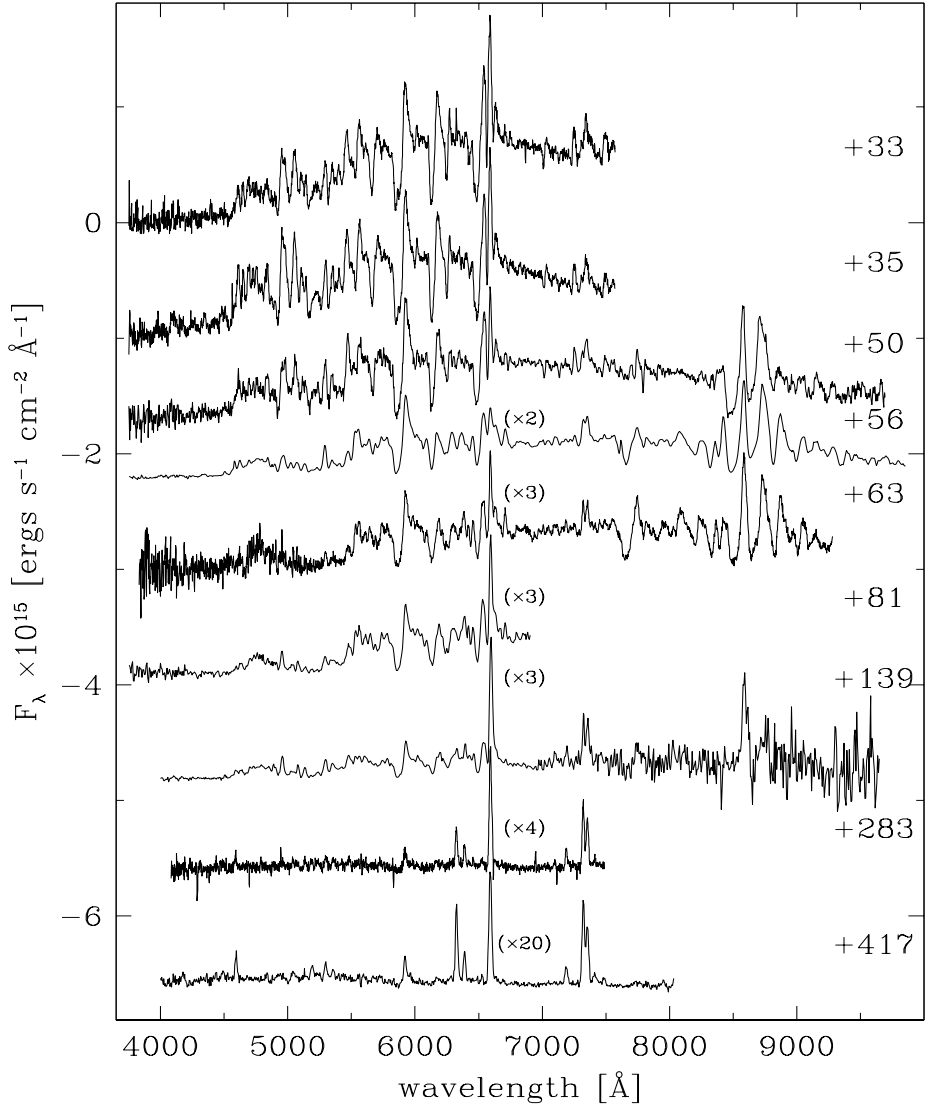,width=13cm,height=16cm,angle=0}
\caption{Spectral evolution of SN~1997D. Wavelength is in the
observer rest frame. The ordinate refers to the first spectrum (+33d),
the other spectra are shifted downwards by $1\times 10^{-15}$,
$1.7\times 10^{-15}$, $2\times 10^{-15}$, $3\times 10^{-15}$,
$3.9\times 10^{-15}$, $4.2\times 10^{-15}$, $5.6\times 10^{-15}$ and
$6.6\times 10^{-15}$ respectively. For clarity some spectra have been
multiplied by factors given in parenthesis.}
\label{spec_fig}
\end{figure*}

The journal of the spectroscopic observations is given in
Tab.\ref{spec_tab}.
 %
%In PaperI we showed
%the spectral evolution of \d\/ up to 10 months after the discovery and
%compared it with that of \a. 
 %
Figure \ref{spec_fig} illustrates the spectroscopic evolution of
\d\/ for over 1 year after discovery.

\begin{table}
\caption{Spectroscopic observations of SN~1997D} \label{spec_tab}
\begin{tabular}{lrcrrr}
\hline
     Date     & phase$^*$ & inst.$^{**}$  &   range
         & res.\\
              & (days)&       &   (\AA)   &    (\AA)   \\
\hline
     14/1/97  &  +33  &  1.5  & 3700-7570  &     7     \\
     15/1/97  &  +34  &  1.5  & 3700-7570  &     7     \\
     16/1/97  &  +35  &  1.5  & 3700-7570  &     7     \\
     17/1/97  &  +36  &  1.5  & 3700-7570  &     7     \\
     27/1/97  &  +46  &  CT4IR& 9645-23180 &    20      \\
     31/1/97  &  +50  &CT1.5  & 3450-9700  &           \\
     6/2/97   &  +56  &  3.6  & 3730-9850  &    17     \\
  12-14/2/97  &  +63  &  2.2  & 3830-9280  &    10     \\
     3/3/97   &  +81  &  3.6  & 3700-6900  &    17     \\
  29/4-1/5/97 &  +139 &3.6+1.5& 3150-10750 &  17+7    \\
    21/9/97   &  +283 &  2.2  & 4100-7480  &    10     \\
    2/2/98    &  +417 &  EF2  & 3860-8030  &    13     \\
\hline
\end{tabular}

$^*$ relative to the estimated epoch of maximum, JD=2450430\\ $^{**}$ See
coding of Table~1, plus 1.5 = ESO 1.5m + B\&C; CT4IR = CTIO 4.0m + IR
spect.; CT1.5 = CTIO 1.5m + R-C spect.
\end{table}

\begin{figure}
\psfig{figure=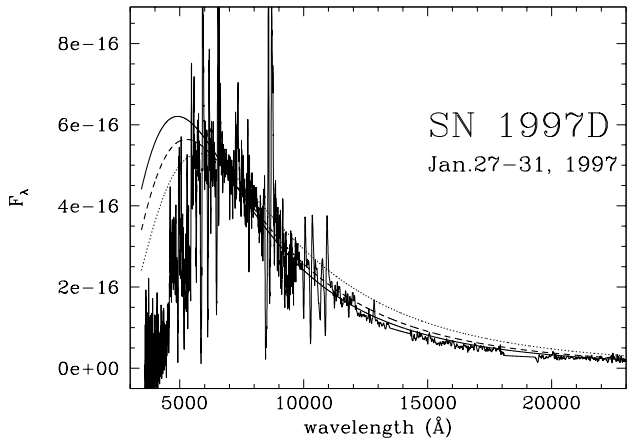,width=9cm,angle=-90}
\caption{Spectral energy distribution between 3500\AA\/ and 2.3$\mu$.
The energy distribution of a black body of 5900, 5500 and 5000\degr~K are 
plotted for reference as a continuous,
dashed and dotted line, respectively.
The IR spectrum is from Clocchiatti et al. (2000)}
\label{sp_uvoir}
\end{figure}

The photospheric spectra are dominated by a red continuum and P-Cygni
profiles of HI, BaII, CaII, NaI, and ScII (see Fig. 4 of PaperI).
Spectral modeling showed that the unusual strength of the BaII lines
compared with typical SNII is due to the low temperature rather than to
overabundance. However, the most striking property of these spectra is the
very low expansion velocity of the ejecta. The minima of the absorption
lines give an expansion velocity of about $1100-1200$ \kms\/ for \Ha\/ and
somewhat larger ($1500-1800$ \kms) for BaII. Spectral modeling suggests
photospheric velocities between 900 \cite{ChuUt97D} and 970 \kms (PaperI)
and kinetic energies between $1-4 \times 10^{50}$ erg, depending on the
ejected mass.  The spectral energy distribution in the optical--IR about
50 days past discovery is shown in Fig.~\ref{sp_uvoir}. As mentioned
above, the distribution is peaked in the optical window and very little
flux is emitted at wavelength smaller than 5000 \AA.

Two months later the \Ha\/ emission starts to dominate over the
BaII 6497 line and becomes the more intense spectral line along
with NaI D and CaII IR lines. Starting 283 days past maximum the
spectrum is that of a typical SNII in the nebular phase but for the
very low kinetic energy.

\subsection{Nebular line identifications} \label{iden}

\begin{figure}
\psfig{figure=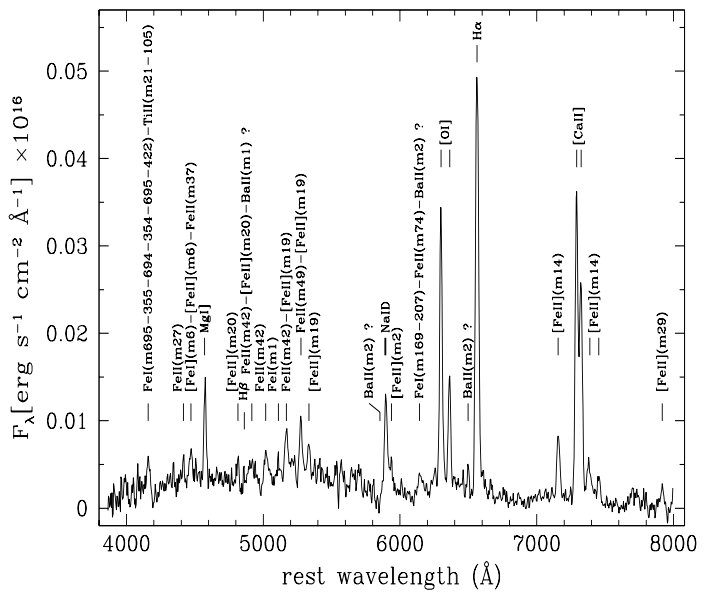,width=10cm,height=9cm,angle=270}
\caption{Line identifications on +417d spectrum. The wavelength is in the
parent galaxy rest frame.}\label{lines}
\end{figure}

The low expansion velocity of the ejecta of \d\/ offers a unique
opportunity to resolve and identify spectral lines.
% never detected before in SNII spectra. 
These identifications are presented in Fig. \ref{lines} for the
spectrum at +417d.

The strongest feature after the prominent \Ha\/ is the 7300 \AA\/
doublet.  Because of the close wavelength coincidence we associate
this emission with the [CaII] lines 7291-7323 \AA\/ and
definitely exclude the contribution of [OII] 7320-7330 \AA\/ doublet.
We remind that the latter was instead identified in the late (about 15yr)
spectra of SNe 1979C and 1980K \cite{fesen}. Just blueward of
this feature an unblended emission is identified with [FeII] 7155
\AA. Though sometimes questioned, this line was already identified in
other SNII at late epochs (e.g. 1988A and 1988H, Turatto et
al. 1993). Here for the first time we single out two other
strong lines of the same [FeII] multiplet 14
($\lambda\lambda$~7388 and 7453 \AA), confirming this
identification. Many other lines of [FeII] and FeII are identified in
the spectrum. In particular, several lines of the FeII multiplet 42
and the [FeII] multiplet 19 are clearly visible.

Another strong line is the emission due to [MgI] 4571 \AA. In
contrast, \Hb \ is very faint if present at all, making the Balmer
decrement $\ge 45$ which indicates collisional excitation.

We could not find an unambiguous identification for a faint line
measured at 6499 \AA \ (rest frame).  For continuity with the
photospheric spectra, we tentatively identify this line with BaII 6497
(mult. 2). To support this identification, we searched the spectrum
for other BaII lines. It turns out that the strongest features
(mult. 1 and 2) if present, would be blended with intense iron
emissions, and therefore we cannot make any definite statement.

More interesting is to analyze in detail the evolution of the [OI]
6300-6364 \AA \ doublet. Spyromilio \& Pinto \shortcite{jasone} have
shown how it is possible to derive an estimate of the OI density from
the evolution of the intensity ratio of the two lines, which ranges
between $6300/6364 =0.95$ when the nebula is optical thick to 3.06 in
case the nebula is optically thin.  In the three nebular spectra of
SN~1997D, we measure $6300/6364 = 0.9 \pm 0.2$ at +139d, $1.9 \pm 0.1$
at +283d and $ 2.65 \pm 0.05$ at +417d. Following Spyromilio \& Pinto
\shortcite{jasone} we can estimate the oxygen number density at day
+283 of $1.5\pm 0.3\times 10^9 {\rm cm}^{-3}$. Noting that the FWHM of
the lines (corrected for instrumental resolution) is 680 km s$^{-1}$
and assuming uniform distribution, we estimate that the oxygen mass in
the ejecta is $\sim 1 M_\odot$. If the filling factor is less than
unity (Spyromilio \& Pinto 1991 estimate a filling factor of $\sim
0.1$ for SN~1987A), the oxygen mass may be even lower. Such an
estimate seems more compatible with a low mass progenitor (see Chugai
\& Utrobin 2000) than with a high mass one, where the total mass of ejected
oxygen is likely to be closer to $2 M_\odot$ (Balberg et al.~2000),
but the latter cannot be ruled out in view of the large observational
uncertainties involved in estimating the oxygen mass.  Furthermore,
the amount of oxygen left in the expanding envelope in the high mass
progenitor case is also sensitive to the extent of fallback on to the
nascent black hole, so a low mass of ejected oxygen does not
necessarily reflect the amount of oxygen in the progenitor.

\section{The Nature of the Progenitor and the Compact Remnant}

In principle, the low luminosity of SN~1997D may be due to either
a low \ni\/ mass or a low opacity to the $\gamma$-rays from 
radioactive decay.
The fact that, for almost one year, the decline rate of the late
time light curve of \d\/ matches the decay rate of \co \ is a strong
indication that the envelope maintains thick to $\gamma$-rays and
hence that the ejected \ni\/ mass is very low (0.002\M). To date
this is the lowest measured value of the \ni\/ mass ejected in a
supernova. The very low expansion velocity of the ejecta and the low
explosion energy of a few $10^{50}$ erg are confirmed by the spectrum
taken 1.5yr after the burst.

These observations single out \d\ as an extremely unusual type II
supernova. As mentioned in the introduction, there are two distinct
types of progenitors that can produce such an explosion - either
a relatively high mass star (26
\M\/; PaperI) or a relatively low mass one (8--10 \M\/; Chugai \& Utrobin
2000). The two models differ in their prediction regarding the nature
of the compact remnant. For the massive star model, significant
fallback of matter is expected because of the very low explosion
energy and the deep gravitational potential of such a massive star. In
contrast, for the 8--12 $M_\odot$ model, the mantle of heavy elements
and the surrounding He/H envelope are so extended that fallback is
very small, even for explosions as weak as $\sim 10^{49}$ erg (Nomoto
et al.~1982; Nomoto 1984, 1987; Nomoto \& Hashimoto 1988). The
significant amount of fallback in the high mass progenitor will most
likely produce a black hole, while in the low mass scenario fallback
is negligible and a neutron star is probably formed. As suggested by
Zampieri et al. (1998a) and recently supported by Balberg et
al. (2000), this distinction is of special interest in the case of \d \
and in the high mass progenitor scenario, since it may offer the first
opportunity of a {\it direct} observation of a newly formed black hole
in a supernova remnant.

\subsection{Light Curve and Spectral Simulations} \label{model}

We performed theoretical calculations of the late time light curve of \d\/ for
both high mass and low mass progenitor models. The mass of the ejecta and the
explosion energy of the high mass progenitor model (constructed from the 25
\ms\ model of Umeda, Nomoto \& Nakamura 2000) are $M_{\rm ej}$ = 18 \ms\ and
$E$ = 4 \e{50} erg. The low mass model has $M_{\rm ej}$ = 6 \ms\ and $E$ = 1
\e{50} erg, as adopted by Chugai \& Utrobin (2000). All the calculations were
performed with a spherically symmetric radiative transfer code, which solves a
$\gamma$-ray transfer equation with one-energy group approximation as well as a
photon transfer equation \cite{iwam00}.

In both models, $^{56}$Ni is distributed uniformly up to an expansion
velocity of 800 \kms, as suggested by the spectral analysis
(PaperI). We find that, in the high and low mass progenitor models,
99\% and 94 \% of the $\gamma$-rays are trapped at day $\sim 400$,
respectively, and thus the optical light curves decline at a rate
close to that of $^{56}$Co, consistent with observations. At day $\sim 400$,
the difference between the two models is too small to be detectable
(the high mass model is only 5\% more luminous than the low mass
model). Significant deviations from the $^{56}$Co decay rate and
differences between the two models ($\sim$ 0.2 mag) are predicted to
be seen only after day $\sim$ 800.

Models suggest that after the end of the plateau, which is the
consequence of an inward-moving recombination wave in the H-envelope,
the light curve of a SNII should be determined mostly by $\gamma$-ray
deposition followed by immediate radiation of optical
photons. Therefore the \co\/ rate should be a lower limit to the slope
of the light curve. However, observations covering the period 60--200
days show that the decline rate was slightly slower than the \co\/
decay rate. This suggest that $\gamma$-ray deposition followed by
instantaneous emission of optical photons might not be the only source
of luminosity for \d\/ at this epoch. This probably means that the
photosphere has not disappeared yet.  Indeed, spectra taken at 81d and
139d (Fig. 3) show the persistence of P-Cygni absorption profiles in
many of the strongest lines (NaI D, BaII, \Ha), even though the CaII
IR triplet is dominated by emission. Note also that the emission
component in
\Ha\ is much stronger than the absorption component, while they are
equivalent in NaI D. Although the contribution of net emission is
significant, a thermal continuum might still influence the spectral
shape.  In addition we find that the B band, where there are no strong
emission lines, declines much more slowly than the redder bands, all
of which include some strong emission lines.

The presence of a photosphere at such advanced phases (its velocity, as
measured from the P-Cygni profile of \Ha, is $\sim1090$ \kms)  might
suggest the presence of an inner region with a steep density gradient,
situated below the H-envelope. This fact might favor a high mass model and
deserves further quantitative study. The spectra taken at late times (283
days and 417days) show only emission lines and, in that period, the light
curve decline rate becomes closer to the \co \ decay. 

We tried to simulate the nebular spectrum of SN~1997D using a NLTE
spectrum synthesis code that computes $\gamma$-ray deposition and heating,
followed by cooling by forbidden line emission. The code is designed for
Type Ia SNe, so H is not included. Fitting the observed FeII forbidden
lines provides support to the finding that the Ni mass is of the order of
0.002 \ms. Also, the velocity of the nebular lines is very small, about
800 \kms, confirming that the kinetic energy of the ejecta is low and that
the small Ni mass is not mixed out significantly. 

\subsection{Black Hole Remnant}

We now turn to discuss the implications of the late times measurements on
identifying the nature of the compact remnant. The data indicate that the
late time bolometric light curve is consistent with the decay of
$^{56}$Co, and there is no evidence up to 400 days of a change in the
decline rate caused by emission from the newly formed compact object. The
bolometric luminosity at $t = 417$ days is $7.8\times 10^{38}$ erg
s$^{-1}$, sufficiently low that emission from a solar mass compact object
accreting at the Eddington limit ($L_{Edd}\approx 2.6 \times
10^{38}\;M_*\;$erg s$^{-1}$, where $M_*$ is the mass of the compact object
in \M\/ and the accreting material is assumed to be hydrogen poor) should
have already been detected through a deviation from a perfect $^{56}$Co
decay. Furthermore, the upper limit in the R band at $\sim700$ days rules
out emission at a luminosity level of $\sim 10^{38}$ erg s$^{-1}$ from the
compact object at that time, and further strengthens the conclusion that
Eddington-limited accretion is not taking place. Therefore, quite
independently of detailed theoretical modeling, these observations in
themselves tend to favor the BH formation scenario.

The extremely low efficiency of converting spherical accretion onto a
black hole into radiative luminosity (for the accretion rates of interest)
severely limits its ability to radiate at the observed rate.
Hence, the absence of a signature from the compact object is
qualitatively consistent with the presence of a black hole in SN~1997D.
In fact, emergence of a black hole as early as $t=417\;$days
would require a very high accretion rate over a relatively
long time. The luminosity from steady-state, hypercritical spherical
accretion of helium-rich matter onto a black hole can be approximately
expressed as
\cite{Blondin86}:
 \begin{equation}\label{eq:Blond} L_{acc}\simeq
 1.44\times 10^{38}\left(\frac{M_{BH}}{M_\odot}\right)^{-1/6}
 \left(\frac{\dot M}{M_\odot\;{\rm yr}^{-1}}\right)^{5/6} {\rm erg} 
 \ {\rm s}^{-1}\,
 \end{equation}
so that to sustain a luminosity of a few $\times 10^{38}$
erg s$^{-1}$ at $t=417\;$days, the accretion rate would have to be a 
few $M_\odot\;{\rm yr}^{-1}$ (depending on the exact 
mass of the black hole). Since the
accretion rate is likely to be declining, it would have had to be at least as 
large prior to $t=417\;$days, implying that several $M_\odot$ of 
material had been accreted in late-time fallback. 
For comparison, the best fit model
for SN~1997D of PaperI has only $0.2\;M_\odot$ of initially bound material
available for late-time fallback (T.\ Young 1999, private communication).
Indeed, the estimated accretion luminosity for $3\;M_\odot$ black hole in
SN~1997D at $t\sim$ hundreds of days is a few times $10^{36}\;$erg s$^{-1}$
(Zampieri et al. 1998a; Balberg et al. 2000), and emergence of the accretion
luminosity is expected only at $t\!\approx\!1000\;$days after the explosion.

We note that if the bound material has sufficient angular momentum,
accretion should proceed via a disk, rather than through a spherical
inflow. The disk would have to be very optically thick and advection
dominated in order to accommodate the hypercritical accretion rates
expected during the first years after the explosion (see Mineshige et al.
1997 for an hydrodynamic study of the properties of such disks up to
$\sim$ 1 day after the explosion).
While the emission properties of such advection-dominated, optically
thick disks at hypercritical accretion rates are poorly understood at
present, disk accretion onto a black hole can, in principle, also be
consistent with the current observations if the accretion efficiency
is low.

\subsection{Neutron Star remnant}

The possibility that the remnant of SN~1997D is a neutron star
can be examined in a similar fashion as the well studied case of SN~1987A. 
To date, the SN~1987A remnant has failed to provide any evidence for the 
presence of a neutron star formed in the explosion.

First, we note that the luminosity at $t\!=\!417\;$days appears to rule out the
formation of a Crab-like pulsar in SN~1997D. The present magnetic-dipole
emission estimated for the Crab pulsar is $\sim 5\times 10^{38}\;$erg s$^{-1}$,
and at an age of 417 days it was probably $\sim 10^{40}\;$erg s$^{-1}$ (Kumagai
et al.~1989; Woosley et al.~1989). If a sizable fraction of such emission is
deposited in the supernova ejecta, it should have had an observable effect on
the luminosity.

If \d\ did produce a neutron star which is weakly magnetized or/and slowly
rotating, this may have been observable if it is accreting material through
late-time fallback.  Initially, the accretion rate is so high that such
fallback is expected to be neutrino-cooled and to produce very little
observable radiative luminosity, but this luminosity should gradually increase
as the expansion of the ejecta causes a decline of the accretion rate
\cite{Chevalier89}.

In the case of SN~1987A, Houck \& Chevalier (1991) found that the radiative
luminosity should reach the Eddington limit about
six months after the explosion, and remains at that value for several years.
Following their approach we can make a similar estimate for accretion onto
a putative neutron star in SN~1997D. At late times the accretion rate is
expected to be ballistic (dust-like) with a secular time dependence of
(Chevalier 1989; Colpi et al.~1996)
 \begin{eqnarray}
 \dot M(t>t_{late}) & \approx & 1.2 \times 10^{-4}
 \left(\frac{M_*}{1.4} \right)
 \left(\frac{\rho_0 t_0^3}{10^9 {\rm g cm^{-3}
 s^3}}\right) \times \nonumber \\
 & & \left(\frac{t_{late}}{10\; {\rm days}}\right)^{-1/3}
 \left(\frac{t}{{\rm yr}}\right)^{-5/3}
 M_\odot {\rm yr}^{-1}\;.
 \label{mdotlate}
 \end{eqnarray}
Here $\rho_0$ is the average initial density of the accreting material
and $t_0$ is the initial expansion time $t_0=r/v$ ($r$ is the radius and
$v$ the velocity of a gaseous shell), assuming initial homologous
expansion at the onset of dust-like flow. $t_{late}$ denotes the time at
which the accretion flow settles into dust-like motion, and is roughly
$t_{late}\approx10\;$days for SN~1987A. Both the numerical coefficient and
$t_{late}$ are weakly dependent on the details of the explosion. The
quantity $\rho_0 t_0^3$ is related to the kinetic energy per unit mass
${\cal E}$ and the amount of mass $M_{bound}$ which is gravitationally
bound to the neutron star at the onset of homologous expansion: $\rho_0
t_0^3 = 0.3^{3/2} (3/4\pi) M_{bound} {\cal E}^{-3/2}$.

The luminosity is expected to reach the Eddington limit roughly when the
accretion shock approaches the trapping radius in the accretion flow. This
corresponds to an accretion rate of $\sim 4\times 10^{-4} M_\odot/{\rm
yr}$ \cite{HouChev91}. Again, this value depends only weakly on the
details of the explosion. Setting ${\dot M} \sim 4\times 10^{-4}
M_\odot/{\rm yr}$ in equation~(\ref{mdotlate}), we find that the
Eddington-limited accretion phase is reached at time
 \begin{equation}
 t_* = 0.5
 \left(\frac{M_*}{1.4} \right)^{3/5}
 \left( \frac{t_{late}}{10 \ {\rm days}} \right)^{-1/5}
 \left( \frac{\rho_0 t_0^3}{10^9 {\rm g cm^{-3} s^3}} \right)^{3/5}
 \, {\rm yr} \,.
 \label{teme}   
 \end{equation}
 Unlike typical Type II supernovae which have a larger $^{56}$Co
abundance, in \d \ this abundance is so low that the radioactive
luminosity is comparable to the Eddington limit for a solar mass compact
object as early as a few hundred days after the explosion. Hence, in \d,
$t_*$ provides a reasonable estimate as to when the fallback luminosity
will ``emerge'' above the emission of radioactive elements, causing the
light curve to deviate from pure $^{56}$Co decay.

Assuming again initial homologous expansion and uniform density throughout
the post-explosion envelope, the amount of mass gravitationally bound to
the neutron star can be roughly approximated as
 \begin{eqnarray}
 M_{bound} & = & 0.013
 \left(\frac{M_*}{1.4} \right)
 \left( \frac{M}{10 M_\odot} \right)^{1/3}
 \left( \frac{r_0}{10^{13} {\rm cm}} \right)^{-1} \times \nonumber \\
 & & \left( \frac{\rho_0 t_0^3}{10^9 {\rm g cm^{-3} s^3}} \right)^{2/3}
 M_\odot
 \, ,
 \label{mbound}
 \end{eqnarray}
 where $M$ and $r_0$ are the mass and radius of the envelope at the onset
of expansion. Equation~(\ref{mbound}) shows that, even in extended
progenitors (outer radius of several $10^{13}$ cm), a small amount of
material, of the order of $10^{-2}M_\odot$, can remain bound to the
neutron star after the explosion and therefore be available for late time
fallback. In fact, $M_{bound}$ increases with increasing $\rho_0 t_0^3$,
i.e. with decreasing explosion (and kinetic) energy. This small amount of
gas is dynamically negligible if compared to the mass of the neutron star.
However, as shown by equation~(\ref{mdotlate}), it is sufficient to
establish a considerable accretion rate at late times ($\sim 10^{-4}
M_\odot$ yr$^{-1}$ at $t\sim 1$ yr) and to give rise to an
Eddington-limited luminosity $\sim$ 1 year after the explosion
(eq.~[\ref{teme}]).

The post-explosion profile of the bound material in the 26 $M_\odot$
progenitor model of PaperI has $\rho_0 t_0^3 \simeq 1.5 \times 10^{10}
{\rm g \ cm^{-3} \ s^3}$, for which the accretion rate would decline to
$4\times 10^{-4} M_\odot/{\rm yr}$ only at $\sim 2.5$ years after the
explosion. Although this model involves forming a compact remnant with a
mass $M \ga 3 M_\odot$ and therefore likely to be black hole, its
parameters do allow for an accreting neutron star to remain hidden in its
midst up to $t=900\;$days. On the other hand, the corresponding parameters
for the low-mass progenitor of Chugai \& Utrobin (2000), which is much
more likely to have produced a neutron star, \ give $\rho_0 t_0^3\simeq
6.3\times 10^{8}{\rm g \ cm^{-3} \ s^3}$. This latter case is more similar
to that of SN~1987A, and accretion onto the neutron star will have reached
the critical rate $4-5$ months after the explosion, and at $t=417\;$days
would be radiating at the Eddington luminosity. The absence of such
luminosity in the light curve of SN~1997D at 400 and $700\;$days could
imply that the remnant is a black hole.

Note, however, that the same reservations originally raised regarding
SN~1987A apply here as well. First, accretion luminosity would not be
produced if the accretion flow is disrupted owing to dynamical
instabilities. Fryer, Colgate \& Pinto (1999) have recently speculated
that such an instability could arise for an accreting neutron star when
the heavier elements begin to recombine, giving rise to a line-driven wind
that eventually expels all the material that was initially bound.
Furthermore, as noted by Mineshige, Nomoto, \& Shigeyama (1993), if an
accretion disk forms, the timescale for the accretion luminosity to
decrease below the Eddington limit depends on the viscosity parameter, and
further study is needed to investigate the evolution in this case.

\section{Conclusions and Discussion}

The analysis of the late-time photometric and spectroscopic data presented
here highlight the peculiar character and unusual properties of
\d\/. The very low expansion velocity of the ejecta ($\sim 1000$ km
s$^{-1}$) and the very low abundance of $^{56}$Ni (0.002 \M\/) make
SN~1997D a unique supernova.  The observations made until 400 days
after the explosion (and an upper limit of the luminosity at 700 days)
lend further support to the interpretation that
this supernova was the result of an unusually low energy explosion
($\sim 10^{50}$ ergs). The available spectroscopic data show also that
at $t=417$ days the supernova ejecta have not yet started to interact
with the CSM.

The observed light curve of \d\/ after day 300 is consistent with pure
radioactive heating, and there is no evidence of a significant
deviation caused by other energy sources in the expanding supernova
envelope. A neutron star formed from the collapse of a low mass
progenitor would have produced emission at the Eddington rate if it
were rapidly rotating and strongly magnetized (like the Crab), or if
it were slowly rotating, weakly magnetized and accreting due to late
time fallback. Such emission would have caused the light curve to
deviate from pure radioactive heating decay by $t = 417$
days. Therefore, the presence of a neutron star remnant seems less
likely. On the other hand, a spherically accreting black hole is
indeed expected to be undetectable in comparison to radioactive
heating during the period of observation, and hence it could not have
revealed its presence (a black hole accreting hypercritically through
a disk could also be undetectable if the accretion flow is
advection-dominated). The formation of a BH remnant is thus consistent
with the available data.

Several interesting implications can be drawn from the conclusion that a BH may
be present in SN~1997D. First, the large amount of fallback needed to produce a
remnant of $\sim 3 M_\odot$ would imply that the progenitor was a massive star.
In this respect, it is interesting to investigate how the explosion energy of a
massive star could be reduced to only a few $10^{50}\;$erg.  We note that
Iwamoto et al. (2000) proposed a scenario in which the core of \d\/
had a small angular momentum because the angular momentum of the core was
transferred via magnetic--field interaction to the massive H envelope. In this
scenario the explosion of \d\/ took place as a low--energy SNII instead of as a
hypernova as in the cases of \bw\/ and \ef, which had similar masses.
Furthermore, if such low-energy supernova have a non-negligible rate (but are
only under-detected), they could have a significant influence on the formation
rate and mass distribution of compact objects, and perhaps even on
nucleosynthesis abundances of heavier elements.

Most notably, if the remnant is a black hole, \d\/ may be the first
case where obtaining direct observational evidence of the emergence of
a black hole in the light curve is technologically possible. Such an
observation would confirm the theoretical assessment that black holes
are formed in supernovae (Timmes et al.~1996; Fryer 1999).  As
mentioned above, in the case of \d\/, the luminosity due to late-time
spherical accretion onto the black hole may thus be observable about
three years after the explosion (Zampieri et al. 1998a; Balberg et
al. 2000). The specific power-law time dependence of this luminosity
should allow to distinguish it from any radioactive source and also
from any possible emission caused by circumstellar interaction (which
in any case has not been observed as late as 1.5 yr after the
explosion). We therefore re-emphasize the need for obtaining deep
observations of \d\/ in the very near future.\\
We will focus our attention also on a more recent supernova,
SN~1999eu, which shows many similarities with SN~1997D and for which
we hope to be able to secure very late photometric and spectroscopic
observations.

\bigskip {\bf ACKNOWLEDGMENTS}
I.J Danziger is acknowledged for helpful suggestions and discussions.\\
This work has been supported in part by the Italian Ministry for
University and Scientific Research (MURST) under grant
cofin--98--2.11.03.07 at the University of Padova, by NSF grant AST
96-18524, and NASA grants NAG 5-7152 and NAG 5-8418 at the University of
Illinois at Urbana-Champaign, and by the grant-in-Aid
for Scientific Research (07CE2002, 12640233) of the Ministry of Education,
Science, Culture and Sports in Japan.

\end{document}